\documentstyle[12pt]{article}
\def\laq{\raise 0.4 ex \hbox{$<$}\kern -0.8 em\lower 0.62 ex\hbox{$\sim$}}
\def\gaq{\raise 0.4 ex \hbox{$>$}\kern -0.7 em\lower 0.62 ex\hbox{$\sim$}}
 
\def\beq{\begin{equation}}
\def\eeq{\end{equation}}
\def\bea{\begin{eqnarray}}
\def\eea{\end{eqnarray}}
\def\bq{\begin{quote}}
\def\eq{\end{quote}}

\def\AJ{{\it Ap. J.} }
\def\AJL{{\it Ap. J. Lett.} }

\def\ANJ{{\it Astron. J.} } 

\def\AP{{\it Ann. Phys. (N.Y.)} }

\def\BAPS{{\it Bull. Am. Phys. Soc.} }

\def\CQG{{\it Class. Quantum Gravity} }
\def\FP{{\it Fortschr. Physik} }
\def\GRG{{\it Gen. Rel. Gravitation} }

\def\JHP{{\it J. High Energy Phys.} }

\def\MPL{{\it Mod. Phys. Lett.} }

\def\NAT{{\it Nature} }
\def\NC{{\it Il Nuovo Cimento} }

\def\PL{{\it Phys. Lett.} }
\def\PR{{\it Phys. Rev.} }
\def\PRL{{\it Phys. Rev. Lett.} }

\def\RPP{{\it Rep. Progr. Phys.} }

\def\RMP{{\it Rev. Mod. Phys.} }

\def\gappeq{\mathrel{\rlap {\raise.5ex\hbox{$>$}}
{\lower.5ex\hbox{$\sim$}}}}
 
\def\lappeq{\mathrel{\rlap{\raise.5ex\hbox{$<$}}
{\lower.5ex\hbox{$\sim$}}}}
 
\begin{document}
\pagestyle{empty}
 
\begin{flushright}
{\sc DF/IST-2.99} \\
{\sc NYU-TH/99/05/03} \\
\end{flushright}

\vspace*{0.5cm}
 
\begin{center}
{\large\bf Nonminimal coupling and quintessence }\\
\vspace*{1.0cm}
{\bf Orfeu Bertolami}\footnote{{Permanent address
Departamento de F\'\i sica, Instituto Superior T\'ecnico, Av. Rovisco Pais, 
1096 Lisboa Codex, Portugal. E-mail: \tt orfeu@alfa.ist.utl.pt}}\\
\medskip
{New York University}\\
{Department of Physics}\\
{Andre and Bella Meyer Hall of Physics}\\
{4 Washington Place}\\
{New York, NY 10003-6621, USA}\\
\vspace*{0.5cm}
and \\
\vspace*{0.5cm}
{\bf  P.J. Martins}\\
\medskip
{Departamento de F\'\i sica}\\
{Instituto Superior T\'ecnico, Av. Rovisco Pais}\\
{1096 Lisboa Codex, Portugal}\\
 
\vspace*{1.0cm}
{\bf ABSTRACT} \\ 
\end{center}
\indent
 
\setlength{\baselineskip}{0.6cm}
 
Recent studies of type Ia Supernovae with redshifts up to about $z~\laq~1$
reveal evidence for a cosmic acceleration in the expansion of the Universe.
The most straightforward explanation to account for this 
acceleration is a cosmological constant dominating the recent history of 
our Universe; however, a more interesting suggestion is to consider 
an evolving vacuum 
energy. Several proposals have been put forward along these lines, most 
of them in the context of General Relativity. 
In this work we analyse the conditions under which the dynamics of 
a self-interacting Brans-Dicke field can account for this accelerated 
expansion of the Universe. We show that accelerated expanding solutions 
can be achieved with a quadratic self-coupling of the 
Brans-Dicke field and a negative coupling constant $\omega$.
 
\vfill
\eject

\setcounter{page}{2}
\pagestyle{plain}

\setcounter{equation}{0}
\setlength{\baselineskip}{0.8cm}
 
\section{Introduction}
 
Type Ia Supernovae (SNe Ia) allow us, when used as standard candles, 
to reliably determine the cosmological parameters. Indeed, low redshift 
supernovae can be used to obtain the Hubble constant, $H_0$, 
while supernovae at greater distances allow probing the deceleration 
parameter, $q_{0}$.
Recent analyses of the magnitude-redshift relation of about 50 SNe Ia 
with redshifts greater than $z \ge 0.35$
strongly suggest that we are living in an accelerating, low-matter density 
Universe [1 - 4]. The consistency 
relationship between these cosmological parameters and the luminosity 
distance, $d_{L}$,  of a SNe is given, for $z~\laq~1$, by 
$d_{L} \approx H_{0}^{-1}[z + (1-q_{0})z^{2}/2]$. The results obtained 
by different groups indicate that [1 - 4]
 
\beq
-1~\laq~q_0~<~0~.
\label{1.1}
\eeq 
 
These values seem to favour, in the case of a flat Universe, a rather low
contribution to the energy density from non-relativistic matter, say
$\Omega_{M} \sim 0.3$, but on the other hand a dominant 
positive cosmological constant, $\Omega_{\Lambda} \sim 0.7$, as can be 
inferred from data \cite{Efstathiou} and consistently from the well known
Friedmann model equation
 
\begin{equation}
\label{1.2}
q_{0} = {1\over 2 } (3 \gamma  + 1) \Omega_{M} - \Omega_{\Lambda}~~, 
\end{equation}
 
\noindent
where $\Omega_{M(V)}$ denotes the energy density of matter (vacuum) 
in terms 
of the critical density, $\gamma$ stands for the constant in the matter 
equation of state, $p =  \gamma \rho$, and observationally 
$-1 \le \gamma \le 0$ 
($\gamma = 0$ for  non-relativistic matter, $\gamma = 1/3$ for relativistic 
matter and $\gamma = -1$ for a scalar field fluid dominated by 
its potential).
Even though a dominant cosmological constant corresponding to
$\Omega_{\Lambda} \sim 0.7$ is 
consistent with all observational facts (see \cite{Bertolami1} 
for a list of constraints and \cite{Sahni} for a thorough discussion) it 
implies a quite unnatural
fine tuning of parameters if its origin lies in any known particle physics 
scale (see Refs. \cite{Weinberg} for reviews and Refs. 
\cite{Bertolami1, Bertolami2}  for possible connections with fundamental 
symmetries such as
Lorentz invariance and string theory S-duality). 
 
From Eq. (\ref{1.2}) we see that a negative $q_{0}$ can also arise from 
a negative pressure component or ``dark energy''.
This alternative calls for a scalar field endowed with a potential
which can give rise to a dynamical vacuum energy, 
the so-called ``quintessence''.
Suggestions along these lines were proposed 
long ago, although yielding, in that case, a vanishing  deceleration 
parameter \cite{Bertolami3}. A number of quintessence models have 
been put forward, most of them invoking a scalar field with 
a very shallow potential, which until recently was overdamped 
by the expansion of the Universe, so that its energy density 
was smaller than the radiation energy density at early times 
\cite{Caldwell, Turner, Zlatev}.
For instance, scalar fields
with an exponential type potential can, under conditions, 
render a negative $q_0$ \cite{Peebles, Ferreira}. More involved possibilities 
include the string theory dilaton together with 
gaugino condensation \cite{Binetruy}, an axion with an almost massless
quark \cite{Kim}, the effect of $\it D$-particle
recoil \cite{Ellis}, supersymmetric inflationary models \cite{Rosati}, 
the multidimensional
Einstein-Yang-Mills system \cite{Bento1}, etc. However interesting, most of 
these suggestions necessarily involve a quite severe fine tuning 
of parameters \cite{Kolda}.
This fact calls for constructions that allow for a negative deceleration 
parameter using sources of negative pressure that do not 
require a potential arising from known particle physics scales.
Other difficulties associated with quintessence is that the 
couplings of the 
scalar field to matter can lead to observable long-range forces and time 
variation of fundamental constants of nature \cite{Carroll, Chiba}.
 
In this work we study the possibility of obtaining the required negative 
pressure effects by considering a scalar 
field coupled non-minimally with gravity. It is well known that 
the non-minimal coupling of fields to gravity have non-trivial 
implications on issues such as  
geodesic completeness, stability of the ground state 
(see \cite{Bertolami4} and references therein) and the variation of 
constants of nature.
In  here, we shall consider the self-interacting Brans-Dicke theory as 
a prototype of a non-minimally coupled theory and show that accelerated 
expanding solutions can be obtained and the abovementioned difficulties 
partially evaded. 
Actually, accelerated expanding solutions at present time, 
the so-called scaling attractors, have already been studied in some theories 
of gravity with non-minimal coupling to a scalar field. Indeed, the non-minimal
coupling ${1 \over 2} \epsilon \phi^2 R$ has been considered in 
Ref. \cite{Uzan}
with the conclusion that the scalar field behaves as a barotropic fluid, 
then leading to scaling attractors, only when the constant 
$\epsilon << 1$ for exponential and power-law potentials. In Ref. 
\cite{Amendola} the search of scaling attractors has been extended to
the case of the non-minimal coupling of the form $[1 + \epsilon f(\phi)] R$
and $V(\phi) = A f(\phi)^{M}$, where $f(\phi)$ is a power-law or exponential 
function of the scalar field and $A$ and $M$ are constants, and 
shown that, in the limit where the kinetic term of the scalar field is
dynamically unimportant, the constraint on the time variation of the 
gravitational coupling is quite severe and limits the fraction of energy in 
the scalar field to be at most $4\%$ of the total energy density. We shall see
in the next sections that our results, although obtained from somewhat
different starting assumptions, are consistent with the conclusions of 
Refs. \cite{Uzan, Amendola}, namely that although accelerated expanding
solutions can be obtained in non-minimally coupled gravity theories 
the constraint on the variability of the gravitational coupling is quite 
strong and that it implies a Universe that is considerably 
older than $H_{0}^{-1}$. 
 
Our starting point is the Brans-Dicke (BD) theory \cite{Brans} as this is 
a viable scalar-tensor alternative to General Relativity. In this theory 
spacetime is described by the interplay of the
metric tensor with a scalar field, $\phi$, so that the strength of the 
gravitational coupling to matter is given by $\phi^{-1}$.
The theory is consistent with observational tests which explains the 
renewed interest in its application both in astrophysics 
and cosmology. Moreover, the BD theory is also the model arising 
from string theory at low-energies in the so-called string or Jordan frame 
and from the dimensional reduction of Kaluza-Klein theories. BD theories
and its extensions are known to have relevant cosmological implications 
\cite{La, Bento2, Kolitch}. 
 
In its simplest version, the salient features of BD theory depend 
uniquely on the strength of the dimensionless 
``Dicke coupling constant'', $\omega$, that couples 
the scalar field universally to matter.
However, the cosmological setting arising from the simplest BD scenario is,
as far as power-law solutions are concerned (cf. equations. below), 
inconsistent 
with the cosmic acceleration, unless $|\omega|$ is of order unity.
Since the coupling $\omega$ is observationally constrained by solar-system 
experiments to be $|\omega| > 500$ \cite{Will}, we are led to consider 
a version of 
the BD theory where the BD scalar has a potential. For the latter we choose 
a quadratic self-coupling in the Jordan frame 
as this implies in a minimal change in the 
field equations. Thus, our scenario can be regarded as the next to 
minimal Brans-Dicke model. Furthermore, we show that a negative $\omega$
is required to account for the accelerated expansion of the Universe.
 
In what follows we consider an homogeneous and isotropic 
Friedmann-Robertson-Walker 
Universe where matter (not including the BD scalar field) is 
described as a perfect fluid. The relevant thermodynamical fluid 
variables are the energy density and the isotropic pressure that are 
related by the equation of state specified above.
 
\section{Field Equations and Solutions}

The self-interacting BD field theory of our interest is described by the
following action:
 
\begin{equation}
\label{eq1}
S={1 \over 16 \pi} \int d^4x \sqrt{-g} \left[ \phi R -\omega \frac
{\phi_{,\alpha} \phi_{,}^{\alpha}}{\phi}  + V(\phi) \right] + S_{Matter}~~.
\end{equation}
 
\bigskip
\noindent
where we choose $V(\phi) = V_{0} \phi^2$. As already mentioned in the 
Introduction a BD theory with a vanishing potential 
is incompatible with a negative
$q_0$, at the least for power-law solutions for the scale-factor and scalar 
field specified below, unless $|\omega| = O(1)$ which is inconsistent 
with solar system experiments. However, as we shall see in a while, 
the potential we have introduced for the scalar field 
changes the equations of motion in a 
minimal way. It is worth stressing that introducing a potential does not
involve, in our proposal, any fine tuning of parameters as our scale, 
$V_{0}$, will be fixed by the value of Newton's constant and the age of 
the Universe. 
Furthermore, as far as giving origin to a time variation of the fundamental 
constants, our proposal implies indeed in a sizable time variation 
of the gravitational coupling, even though this is compatible 
with available data. 
 
From the Lagrangian density (\ref{eq1}) we obtain the field equations:
 
\begin{equation}
\label{eq2}
G_{\mu \nu}= \frac{\omega}{\phi^2} \left[ \phi_{;\mu} \phi_{;\nu} -
\frac{1}{2} g_{\mu \nu} \phi_{;\alpha} \phi_{;}^{\alpha} \right] +
\frac{1}{\phi} \left[ \phi_{;\mu} \phi_{;\nu} - g_{\mu \nu}
\Box^2 \phi \right] 
+ \frac{V(\phi)}{2\phi} g_{\mu \nu} +
\frac{8\pi}{\phi} T_{\mu \nu}^{Matter}~~,
\end{equation}
 
\begin{equation}
\label{eq3}
\Box^2  \phi = \frac{8\pi}{3+2\omega} T^{Matter} 
+ \frac{1}{3+2\omega}\left[ 2V(\phi) - \phi
 \frac{dV(\phi)}{d\phi} \right]~~.
\end{equation}
 
\noindent
Hence for our choice of the potential, $V(\phi)$, the 
last term in Eq. (\ref{eq3}) vanishes and $V(\phi)$ affects directly 
only the dynamics of the scale-factor, $a(t)$.
 
These field equations in a Friedmann-Robertson-Walker geometry 
 
\begin{equation}
\label{eq4}
ds^2 = -dt^2 + a(t)^2   \left[ {dr^2 \over 1-kr^2 } + r^2 d{\Omega}^{2}
\right]~~,
\end{equation}
read
 
\begin{equation}
\label{eq7}
\left( \frac{\dot{a}}{a} \right)^2 + \frac{k}{a^2} =
\frac{8\pi}{3\phi}~\rho - \frac{\dot{a}}{a} \frac{\dot{\phi}}{\phi} +
\frac{\omega}{6} \left( \frac{\dot{\phi}}{\phi} \right)^2 -
\frac{V_{0}}{6} \phi~~,
\end{equation}
 
\begin{equation}
\label{eq8}
3 \left( \frac{\ddot{a}}{a}  \right) = - \frac{8\pi}{(3+2\omega)\phi} 
\left[ (3+\omega)\rho + 3\omega p \right] - \omega \left(
\frac{\dot{\phi}}{\phi} \right)^2 + 3 \frac{\dot{a}}{a}
\frac{\dot{\phi}}{\phi} - \frac{V_{0}}{2}\phi~~,
\end{equation}
 
\begin{equation}
\label{eq9}
\ddot{\phi} + 3 \frac{\dot{a}}{a}\dot{\phi} = 
\frac{8\pi}{3+2\omega}(\rho - 3p)~~.
\end{equation}
 
\noindent
Clearly, these equations must be considered together with the energy 
conservation equation
 
\begin{equation}
\label{eq11}
\dot{\rho} + 3 \frac{\dot{a}}{a}(\rho +3p) = 0~~,
\end{equation}
arising from the covariant conservation of the energy-momentum tensor of 
matter that has been assumed to behave as a perfect fluid.
 
Finally, the relationship between 
the gravitational coupling $G(t)$ and the scalar field is given by
 
\begin{equation}
\label{eq13}
G(t)= \left( \frac{2\omega + 4}{ 2\omega +3}  \right) \frac{1}{\phi(t)}~~.
\end{equation}

In order to obtain solutions for the above equations we consider 
the following power-law form for both scale-factor and scalar field:
 
\begin{equation}
\label{eq14}
a(t)=a_{0} \left( \frac{t}{t_{0}}  \right)^{\alpha} \quad ; 
\quad  \phi(t)= \phi_{0} \left(\frac{t}{t_{0}} \right)^{\beta}~~,
\end{equation}
where the zero indices stand for the present time.
 
Substituting these solutions into Eq. (\ref{eq9}) leads, when neglecting
the mattter pressure constribution ($\gamma \sim 0$), to the relationships
 
\begin{equation}
\label{eq15}
3 \alpha = 2 - \beta \quad ; \quad \beta = \frac{\sigma}{3+2\omega}~~,
\end{equation}
where 
 
\begin{equation}
\label{eq16}
\sigma \equiv \frac{8\pi \rho_{0}t_{0}^2}{\phi_{0}}~~.
\end{equation}
 
\noindent
From Eqs. (\ref{eq13}) -- (\ref{eq15}) and 
the Hubble parameter at present, $H_{0} \equiv {\dot{a} \over a}(t_0)$, 
we get the deceleration parameter, 
$q_{0} \equiv - {\ddot{a} a \over \dot{a}^2}(t_0)$, and the time 
variation of the gravitational coupling:  
 
\begin{equation}
\label{eq17}
q_{0}= \frac{1+\beta}{2-\beta}~~,
\end{equation}
 
\begin{equation}
\label{eq18}
\left( \frac{\dot{G}}{G} \right)_{0} = \frac {3\beta}{\beta -2} H_{0}~~.
\end{equation}
 
\noindent
Moreover, adjusting the gravitational coupling to its present value, 
$G_{N}$, i.e. Newton's constant, we get, after inserting 
Eq. (\ref{eq13}) into 
Eq. (\ref{eq16}) and combining with Eq. (\ref{eq15}), for the age of the 
Universe
 
\begin{equation}
\label{eq20}
t_{0}^2 = \frac{2\beta (\omega + 2)}{3 \Omega_{M} H_{0}^2}~~.
\end{equation}
 
\noindent
Naturally, for consistency $V_{0}$ must satisfy Eq. (\ref{eq7}). 
Hence, for a flat Universe ($k=0$) we find
 
\begin{equation}
\label{eq21}
V_{0}= 3G_{N}\Omega_{M}H_{0}^2 (2\omega +3) 
\left[ \frac{2\beta (2\omega +3) -6\alpha (\alpha + \beta) + 
\omega \beta^2}{4\beta (\omega + 2)^2} \right]~~.
\end{equation}

\section{Conditions for negative $q_{0}$}

In order to have $q_{0} <0$ it is equivalent that the right hand side 
of Eq. (\ref{eq8}) is positive. It is useful then to consider it as a function
 
\begin{equation}
\label{eq22}
f \equiv - \frac{8\pi}{(3+2\omega)\phi} \left[ (3+\omega) \rho \right]
-\omega \left( \frac{\dot{\phi}}{\phi} \right)^2 + 3 \frac{\dot{a}}{a}
\frac{\dot{\phi}}{\phi} - \frac{V_{o}}{2} \phi
\end{equation}
 
\noindent
and establish the conditions for having $f > 0$.
 
\noindent
Power-law solutions Eq. (\ref{eq14}) satisfy the field equations only if 
$\beta = -2$. Then, solving for $\omega$ yields
 
\begin{equation}
\label{eq23}
\omega < - \frac{V_{0} \phi_{0} t_{0}^2}{4} -1~~.
\end{equation}
 
\noindent
Since the product $V_{0}\phi_{0}t_{0}^2$ is completely determined by Eqs. 
(\ref{eq13}), (\ref{eq20}) and (\ref{eq21}) it is easy to see that 
this condition is satisfied for any value of $\omega$. Therefore, we 
can conclude that the cosmic accelerated expansion can be driven by 
a self-interacting Brans-Dicke field as we have specified.

In summary we have, after establishing that $\beta = -2~$
 
\begin{equation}
\label{eq24}
a(t)=a_{0} \left( \frac{t}{t_{0}} \right)^{\frac{4}{3}} \quad ; \quad
\phi (t)=\phi_{0} \left( \frac{t}{t_{0}} \right)^{-2}~~,
\end{equation}
 
\begin{equation}
\label{eq25}
t_{0}^2 = - \frac{4(\omega + 2)}{3\Omega_{M}H_{0}^2}~~,
\end{equation}
 
\begin{equation}
\label{eq26}
\left(\frac{\dot{G}}{G} \right)_{0} = \frac{3}{2} H_{0}~~, 
\end{equation}
 
and
\begin{equation}
\label{eq27}
q_{0}= - \frac{1}{4}.
\end{equation}
 
\bigskip
\noindent
We see that the obtained value for $q_{0}$ is consistent 
with the observations, Eq. (\ref{1.1}), and so is 
the resulting time variation of the gravitational coupling 
\cite{Will}, even though many searches are consistent with no variation 
at all \cite{Gillies}. We stress that, in our proposal, 
the gravitational coupling is a growing function of time. Morever, 
notice that, although $f$ is positive 
independently of the sign of $\omega$, as $\beta$ is negative then in 
order to have a meaningful age of the Universe it implies that the 
Dicke coupling constant must be negative.
Interestingly, negative values for $\omega$ are found 
in the BD effective low-energy models arising from Kaluza-Klein 
and superstring theories \cite{Kolitch}. We shall see in the next section
that negative values for $\omega$ are also required for obtaining growing 
modes for the energy density perturbations in a Universe expanding in
an accelerated way.
 
 
\section{Perturbed Field Equations and Asymptotic Behaviour}
 
Having found accelerated expanding solutions in the context of the
BD theory we should ask whether the issue of structure formation is 
modified by the dynamics of the BD scalar field.
That is, we are bound to consider the evolution of the energy density 
perturbations in the context of our BD scenario as these are, of course, 
associated with the formation of structure in the Universe. 
Hence in what follows we shall develop a formalism that quickly allows
us to find the asymptotic behaviour of the relevant variables for accelerating
expanding solutions at present time. This treatment is similar, in its results,
to the more encompassing analysis of cosmological perturbations for 
generalized theories of gravity \cite{Hwang}, at least in what concerns the
asymptotic behaviour of solutions. 
 
In order to get the matter energy density perturbations 
we consider the temporal components of Eqs. (\ref{eq8}), 
(\ref{eq9}) and (\ref{eq11}) after perturbation.
Then to obtain the relevant perturbed equations we write the metric 
tensor as
 
\begin{equation}
\label{eq28}
g_{\mu \nu} \rightarrow g_{\mu \nu} + \delta g_{\mu \nu}~~,
\end{equation}
 
\bigskip
\noindent
where $\delta g_{\mu \nu} = h_{\mu \nu}$~, and work in the gauge $h_{0
\mu} =0$. Our conclusions will be independent of this gauge choice (see 
\cite{Hwang} for a discussion). For this perturbed metric \cite{Baptista}
  
\begin{equation}
\label{eq29}
\delta R_{00} = \frac{1}{a^2} \left[ \ddot{h}_{kk} -2 \frac{\dot{a}}{a}
\dot{h}_{kk} +2 \left[ \left( \frac{\dot{a}}{a} \right)^2 -
\frac{\ddot{a}}{a} \right] h_{kk} \right]~~.
\end{equation}
 
\noindent
The perturbation of the energy-momentum tensor is given by
 
\begin{equation}
\label{eq30}
\delta T^{00} = \delta \rho~~,
\end{equation}
 
\noindent
and the corresponding trace reads 
 
\begin{equation}
\label{eq31}
\delta T = \delta \rho -3\delta p~~.
\end{equation}
 
\noindent
For the perturbation of the d'Alembertian of the BD field we have
 
\begin{equation}
\label{eq32}
\delta \Box^2  \phi = \delta \ddot{\phi} + 
a\dot{a}h^{kk}\dot{\phi} -
\frac{1}{2a^2} \dot{h}_{kk} \dot{\phi} + 3 \frac{a}{a} \delta \dot{\phi} -
\frac{\nabla^2}{a^2} \delta \phi~~.
\end{equation}
 
\noindent
The relevant perturbations are parametrized in the following way:
 
\begin{equation}
\label{eq33}
h_{kk}=a^2 h~~;
\end{equation}
 
\begin{equation}
\label{eq34}
\delta \phi = \lambda \phi \quad ,  \quad \lambda << 1~~;
\end{equation}
 
\begin{equation}
\label{eq35}
\delta \rho = \Delta \rho \quad , \quad \Delta << 1~~,
\end{equation}
where $h(t)$, $\lambda (t)$ and $\Delta (t)$ are the perturbed 
gravitational field, scalar field and matter energy density, respectively.
 
Since structure is formed when pressure no longer prevents gravitational 
collapse we set $\gamma =0$~. The perturbed equations are then
 
\begin{equation}
\label{equation36}
\frac{1}{2} \ddot{h} + \left( \frac{\dot{a}}{a} \right) \dot{h} =
\frac{8\pi \rho}{\phi}  \frac{(2+\omega)}{(3+2\omega)} (\Delta - \lambda) +
\ddot{\lambda} +2 (1+\omega)  \frac{\dot{\phi}}{\phi} \dot{\lambda} -
\frac{V_{o}}{2} \phi \lambda~~,
\end{equation}
 
\begin{equation}
\label{eq37}
\ddot{\lambda} +  \left( 2 \frac{\dot{\phi}}{\phi}  +
3\frac{\dot{a}}{a} \right) \dot{\lambda} + \left( \frac{\ddot{\phi}}{\phi}
+ 
3 \frac{\dot{a}}{a} \frac{\dot{\phi}}{\phi} \right)  \lambda -\frac{1}{2} 
\frac{\dot{\phi}}{\phi}\dot{h}  - \frac{\nabla ^2 \lambda}{a^2} = \frac{8\pi
\rho}{(3+2\omega) \phi} \Delta~~,
\end{equation}
 
\begin{equation}
\label{eq38}
\dot{\Delta} - \frac{1}{2} \dot{h} + \delta U^{k}_{,k} = 0~~,
\end{equation}
where $U_\mu$ is the comoving fluid velocity.
 
\noindent
Inserting solutions (\ref{eq14}), (\ref{eq15}) with $\beta = - 2$ 
into this set of coupled differential equations we get
 
\begin{equation}
\label{eq39}
\frac{1}{2} \ddot{h} + \frac{\alpha}{t} \dot{h} = \beta (2+\omega)
\frac{(\Delta - \lambda)}{t^2} + \ddot{\lambda} + 2
(1+\omega) \frac{\beta}{t} \dot{\lambda} - \frac{V_{0}}{2} \phi_{0}
\left(\frac{t}{t_0}\right)^{\beta} \lambda~~,
\end{equation}
 
\begin{equation}
\label{eq40}
\ddot{\lambda} +  \left( \frac{\beta +2}{t} \right) \dot{\lambda} + 
\frac{\beta}{t^2} \lambda  - \frac{1}{2} \frac{\beta}{t} \dot{h} -
\frac{\nabla^2 \lambda}{a^2} = \beta \frac{\Delta}{t^2}~~,
\end{equation}
 
\begin{equation}
\label{eq41}
\dot{\Delta} - \frac{1}{2} \dot{h} + \delta U^{k}_{,k} = 0~~.
\end{equation}
 
\noindent
In order to continue we suppose that the perturbations behave as plane 
waves:
 
\begin{equation}
\label{eq42}
\lambda(\vec{r}, t) = \lambda (t) exp (-i \vec k \cdot \vec{r})~~,
\end{equation}
where $k$ is the wave number of the perturbation, and we 
set $\delta U^{k}$ to vanish, which is allowed by an infinitesimal 
gauge transformation.
 
\noindent
Combining then Eq. (\ref{eq39}) with (\ref{eq41}) we get
 
\begin{equation}
\label{eq43}
\ddot{\Delta} + \left( \frac{14}{3t}\right) (\dot{\Delta} - \dot{\lambda}) 
+ \left( \frac{6+2\omega}{t^2} \right) (\Delta - \lambda)  
+ \left( \frac{4+ 4\omega + \frac{14}{3}}{t}\right)  \dot{\lambda} 
+ \frac{V_{0}\phi_{0} t_{0}^2}{2t^2} \lambda 
+ \frac{k^2}{a_{0}^2} \left(\frac{t_{0}}{t}\right)^{\frac{8}{3}} 
\lambda = 0~~.
\end{equation}
 
\noindent
Since we are interested in the asymptotic regime we keep only terms up 
to $t^{-2}$ and neglect the last term of the previous equation.
 
\noindent
Finally, from the value of the product, 
$V_{0}\phi_{0}t_{0}^2 = - (4 \omega + {20 \over 3})$ 
previously obtained, we get the differential equation
 
\begin{equation}
\label{eq44}
\ddot{\Delta} + \frac{C_{1}}{t}(\dot{\Delta} - \dot{\lambda}) +
\frac{C_{2}}{t^2} (\Delta - \lambda) + \frac{C_{3}}{t} \dot{\lambda} +
\frac{C_{4}}{t^2} \lambda = 0~~,
\end{equation}
where $C_{1}= {14 \over 3}$,
 
\begin{equation}
\label{eq46}
C_{2}(\omega) \equiv 6+2\omega~~,
\end{equation}
 
\begin{equation}
\label{eq47}
C_{3}(\omega) \equiv 4(1 +\omega) + C_{1}~~,
\end{equation}
and 
 
\begin{equation}
\label{eq48}
C_{4}(\omega) \equiv - \frac{10 + 6\omega}{3}~~.
\end{equation}
 
\noindent
Aiming to solve Eq. (\ref{eq44}) we look for solutions of 
the following form:
 
\begin{equation}
\label{eq49}
\Delta - \lambda = f(t) \quad ; \quad f(t)=\xi t^{\delta} \quad ; \quad
\Delta (t)= \chi t^{\theta}~~,
\end{equation}
where $\xi$ and $\chi$ are constants.
 
\noindent
Inserting these solutions into Eq. (\ref{eq44}) we obtain the
algebraic equation
 
\begin{equation}
\label{eq50}
\chi \theta^2 + \left[ \chi (C_{3}-1) + \xi (C_{1}-C_{3})  \right] \theta +
C_{4}(\chi -\xi) +\xi C_{2} = 0~~,
\end{equation}
and also that $\delta = \theta$.
 
Solving for $\theta$ we find:
 
\begin{equation}
\label{eq51}
\theta_{\pm} =  \frac{ \chi (1-C_{3}) + \xi (C_{3}-C_{1})  \pm
\sqrt{ \left[ \chi (C_{3}-1) +\xi (C_{1}-C_{3}) \right]^2 - 4\chi \left[
C_{4} (\chi - \xi) + \xi C_{2}  \right] }}{2\chi}~~.
\end{equation}

\noindent
Thus, the existence of growing modes for the energy density perturbations
corresponds now to whether there are
positive $\theta$ solutions. In the case of our interest, namely 
$\omega$ negative and $|\omega | >> 1$, it is easy to see that
 
\begin{equation}
\label{eq52}
\theta_{\pm} \rightarrow  2 \vert \omega \vert \left[ 
\left(1 - {\xi \over \chi}\right) \pm 
\sqrt{\left(1 - {\xi \over \chi}\right)^{2} - 
{1 \over 2 \vert \omega \vert}\left(1 - 2~{\xi \over \chi}\right)}~\right]~~,
\end{equation}
 
\noindent
meaning that $\theta_{+}$ corresponds to a growing mode whenever 
$\xi < \chi$. Moreover, we see that it is 
only for $\omega$ negative that real $\theta$ solutions always exist.
We can then conclude that the cosmological setting of our accelerated 
expanding solutions does not upset known structure formation 
scenarios.
 
\section{Discussion and Conclusions}
 
In this work we have shown that a minimally extended BD theory 
with a quadratic self-coupling in the Jordan frame 
and a negative $\omega$ can account for the accelerated expansion of the
Universe yielding $q_0 = - {1 \over 4}$. The resulting variation on time of 
the gravitational coupling is given by $\left({ \dot{G} \over G}\right)_0 = 
{3 \over 2}~H_0$, being still compatible with data. In this respect our 
results are similar to the conclusions of Refs. \cite{Chiba, Amendola} 
as likewise we find that the existence of scaling attractors in gravity 
theories with non-minimal coupling to a scalar fields is 
severly constrained by the time variability of the gravitational coupling. 
Furthermore, we have also 
shown that the model allows growing modes for the energy density 
perturbations of matter implying that
the dynamics of the BD field does not upset known structure formation 
scenarios .
 
Before closing, it is worth pointing out some distinct features of our 
model. 
The first one being that as accelerated expanding solutions require a 
negative
$\omega$, we then expect the parametrized-post-Newtonian (PPN) parameter 
$\gamma_{PPN} \equiv {1+\omega \over 2+\omega}$ to be, 
for large $|\omega|$, fairly close but greater than 1. This is a clear 
observational signature of our proposal, which even though being 
consistent with current data \cite{Reasenberg}
 
\begin{equation}
\label{eq53}
|\gamma_{PPN} - 1| < 2 \times 10^{-3}~~,
\end{equation}
can be, at least in principle, further improved 
by dedicated solar system experiments. 
 
A second consequence of our
proposal is that it implies the Universe is considerably older than $H_0^{-1}$.
This is incompatible with observation if the age of the 
Universe is identified with age of the oldest stars, the globular clusters.
Contradiction with models of chemical evolution of galaxies may also 
exist. Even though these constraints are extracted from data assuming 
a Universe with constant gravitational coupling they point out that 
the minimal extention of the BD model we have considered is not quite 
consistent with data. In this respect, it is interesting that
the effect of variation on time of the gravitational coupling may reveal 
itself in the evolution of astrophysical objects. Indeed, 
as recently discussed \cite{Torres}, stringent bounds on $\omega$ 
can be set, namely that $\omega > 5000$ and that 
$\left({ \dot{G} \over G}\right)_0 = O(10^{-14})$, based on the 
luminosity function of white dwarfs assuming that $ \dot{G} < 0$ 
and $12.5$ Gyr as the age of the Universe.
We suspect that this age of the Universe problem is a common feature of all 
scalar-tensor gravity models that have scaling attractors and do not
contain, as for instance in \cite{Amendola}, the Einstein-Hilbert term in the
action.

Finally, we could conclude remarking that 
an interesting theoretical challenge would be 
devicing inflationary models which would, at late times, behave like 
scalar-tensor gravity theories in what concerns the existence of 
scaling attractors.
 

{\large\bf Acknowledgments} 

\noindent
One of us (O.B.) would like to thank 
Funda\c c\~ao para a Ci\^encia e a Tecnologia (Portugal) for the sabbatical 
grant BSAB/95/99. 

\newpage


\end{document}